\documentstyle[12pt,epsf]{article}
\newcommand{\BAR}{\overline}

\newcommand{\lsim}{\mathrel{\lower4pt\hbox{$\sim$}}
\hskip-12.5pt\raise1.6pt\hbox{$<$}\;}
\def\uglu{\hskip 0pt plus 1fil minus 1fil}
\def\uglux{\hskip 0pt plus .75fil minus .75fil}
\def\slashed#1{\setbox200=\hbox{$ #1 $}
    \hbox{\box200 \hskip -\wd200 \hbox to \wd200 {\uglu $/$ \uglux}}}

\def\slq{\slashed q}

\newcommand{\ep}{{\eta^\prime}}
\newcommand{\etac}{{\eta_c}}
\newcommand{\te}{\theta_\eta}

\newcommand{\gsim}{\mathrel{\lower4pt\hbox{$\sim$}}
\hskip-12.5pt\raise1.6pt\hbox{$>$}\;}

\begin{document}

\begin{flushright}
JLAB-THY-97-~~~~ \\
BNL-~~~~~~~~~~~~~~~
\end{flushright}
\bigskip
\begin{center}
{\bf 
$B\to \ep+X$ and the QCD Anomaly
}
\\
David Atwood$^a$, 
and Amarjit Soni$^b$
\end{center}

\bigskip

\begin{flushleft}
$a$) Theory Group, CEBAF, Newport News, VA\ \ 23606 \\
$b$) Theory Group, Brookhaven National Laboratory, Upton, NY\ \ 11973
\end{flushleft}
\bigskip

\begin{quote} 
{\bf Abstract}:  Mechanisms for the observed large $Br(B\to \ep +X_s)$
are examined. 
We propose that the  dominant
fraction of the $B\to\ep +X_s$ rate 
is due mainly to $b\to s g^*$, where $g^*$ is an off-shell
gluon, followed by $g^*\to g\eta^\prime$ via the anomalous coupling of the
$\eta^\prime$ to two gluons. The calculated rate for
$B\to\ep+X_s$ is in rough accord with experiment
using
a fairly constant glue-glue-$\ep$ form factor. 
This behavior of the form factor may be indicative 
of glueball dominance of the channel. 
Searches via the modes $\ep h^+h^-$ ($h=\pi$ or $K$)
may be worthwhile.
Charmonia contributions [i.e\ $B\to \eta_c, \psi (\to\ep+X)
+X_s$] can only account for at most 20\% of the 
central value of the signal. 
Implications for $B\to \ep +X_d$ and for the corresponding
$\eta$ modes are also given.
\end{quote} 
\newpage

\section{Introduction}

Recently CLEO has reported\cite{cleo_data} a very large branching ratio for the
inclusive production of $\ep$ (subject to the indicated cut):
\begin{equation}
{\rm Br}( B\to \ep+X; 2.2\leq E_{\ep}\leq 2.7 GeV )= 
(7.5\pm 1.5\pm 1.1) \times 10^{-4}
\label{EQone}
\end{equation}
In this Letter we report on possible mechanisms for this large signal. Two
interesting origins are: (1) $b\to s+g^\ast$ (where $g^\ast$ is an off-shell
gluon) followed by $g^\ast\to\ep +g$ due to the anomalous $\ep$-$g$-$g$ vertex.
(2) $b\to{}$charmonia (e.g.\ $\psi,\etac$)${}+X_s$ 
followed by $\psi(\etac)\to\ep
+X$. We discuss these below in turn.

\section{Role of the Anomaly}

We suggest that the dominant contribution to the observed 
large $B\to\ep+X$ 
signal
is due to
two crucial aspects 
of the Standard Model (SM). First, the SM predicts a very
large branching ratio for the charmless $b\to s$ penguin
transition\cite{hou_soni,Grigjanis}:
\begin{equation}
Br(B\to X_s)\sim 10^{-2}
\label{EQtwo}
\end{equation}
(where $X_s$ is a charmless state containing a $s$-quark). On the quark level
these may be regarded as due to $b\to s g^*$ 
transitions, where $g^*$ is an
off-shell gluon, and it is the gluons from this reaction that could be the
source for the large $\ep$ production. This conversion of gluons to $\ep$
states may be 
accomplished through another important aspect of the SM, namely the
QCD anomaly:
\begin{equation}
\partial_\mu j^\mu_5=
{3\alpha_s\over 4\pi} G_{\mu\nu}\tilde G^{\mu\nu}
+ 2i
\sum_{q=uds} m_q \BAR q \gamma^5 q
\label{EQthree}
\end{equation}
where $j_5^\mu$ is the singlet axial current\cite{schroeder}. Indeed it is the
anomalous contribution to the divergence of the axial current that
distinguishes the SU(3) singlet $\eta_1$ state from the octet pseudo-goldstone
$\eta_8$ state. Intuitively it is the gluonic cloud of the $\ep$ which makes it
appreciably heavier than the members of the  octet: $\pi$, $K$ and
$\eta_8$\cite{derujula}. The physical $\eta$ and $\ep$ states are related to
these according to the mixing
\begin{eqnarray}
\eta &=&\cos\te \eta_8 - \sin\te \eta_1
\nonumber\\
\ep&=&\sin\te \eta_8 + \cos\te \eta_1
\label{EQfour}
\end{eqnarray}
where the mixing angle $\te$ is estimated\cite{pdb} in the range of $-10^\circ$
to $-20^\circ$ so $\ep$ is dominantly $\eta_1$.

In order to model the gluonic coupling of the $\eta_1$ we use an $\ep$-glue-glue
effective vertex: 
\begin{equation}
H(q^2_1, q^2_2, q^2_{\ep}) 
\delta^{ab}
\epsilon_{\mu\nu\alpha\beta}
q^\mu_1 q^\nu_2 \epsilon^\alpha_1 \epsilon^\beta_2
\label{EQfive}
\end{equation}
where $q_1,q_2$ and $\epsilon_1,\epsilon_2$ 
are the 4-momenta and polarizations
of the two gluons and 
$a$, $b$ are
color indices. Thus the picture that we have
for the inclusive $\ep$ production is a combination of $b\to s g^*$ followed by
$g^*\to \ep g$ as shown in Fig.~1. Indeed there have been
many phenomenological attempts to quantify the gluonic content of the
$\ep$\cite{rosner,Tytgat}. In the effective vertex (\ref{EQfive})
$H$ is a form factor that is a general function of the momenta,
$q^2_1,q^2_2$ and $q^2_{\ep}$. If we expand this function in $q^2_1$
and $q^2_2$ at $q^2_{\ep} = m^2_{\ep}$, the leading term is generated by
the QCD anomaly. 
We will determine $H(q^2_1\approx 0, q^2_2\approx 0, m^2_{\ep})$
by studying $\psi\to \ep\gamma$.

As is well known, the $b\to s g^*$ transition 
can be parameterized 
in the standard model 
(in the limit $m_s\to 0$)
by the
induced chromo-electric and chromo-magnetic 
form factors \cite{hou_soni,Grigjanis}:
\begin{eqnarray} 
\Lambda_{e,\mu}^{b\to s} &=&
+
{G_F\over \sqrt 2}
\sum_{i=uct} v_i \BAR s (\lambda^a /2) \left[ F_1^i (q^2) (\gamma_\mu
q^2-q_{\mu}\slq)L \right] b \nonumber\\ 
\Lambda_{m,\mu}^{b\to s} &=& 
- 
{G_F\over \sqrt 2}
{g_sm_b\over 2 \pi^2} \sum_{i=uct} v_i \BAR s 
(\lambda^a /2) \left[ F_2^i (q^2)
i\sigma_{\mu\nu}q_\nu R\right ] b \label{EQsix} 
\end{eqnarray} 
where $v_i=V^*_{is}V_{ib}$, $g_s$ is the QCD coupling constant, 
$q=p_b-p_s$ is the outflowing
gluon momentum, 
$\lambda^a$ are the Gell-Mann matrices, $R=(1+\gamma^5)/2$ and
$L=(1-\gamma^5)/2$.

Using Eq.~\ref{EQsix} and Eq.~\ref{EQfive} we can deduce the differential decay
rate for 
$b(p_b)\to\ep(q_\ep) s(p_s) g(k)$:
\begin{equation}
{d^2\Gamma\over ds dt}
=
{H^2v_t^2G_F^2\over 384 \pi^3 m_b^3}
\left[
\hat F^2(2V-(m_b^2-s)W/2)
+
2\hat G^2(m_b^2-s)V
+
4m_b\hat F \hat G V
\right]
\label{EQseven}
\end{equation}
Here $s=(q_{\ep}+k)^2\equiv q^2$, $t=(p_s+k)^2$ and $u=(p_s+q_{\ep})^2$,
$V=(stu-m_\ep^2 m_b^2 t)/4$, $W=-(s-m_\ep^2)^2/2$, $\hat G=g_sm_b
f_2/(2\pi^2 s)$ and $\hat F=f_1-m_b\hat G$. We have neglected the small
contribution of the $u$-quark and defined $f_i=F^t_i-F^c_i$.

We can now readily calculate the rate for $b\to s g \ep$ if we assume
that $f_i,H$ are roughly constant as a function of $s=q^2$:
\begin{equation}
\Gamma(b\to s \ep g)
= 
{m_b^2 H^2\Gamma_0\over 8}
(
f_i^2\tau_0(x)
+{g_s\over 2\pi^2}f_i f_2\tau_1(x)
+{g_s^2\over 4\pi^4}f_2^2\tau_2(x)
)
\label{EQeight}
\end{equation}
where $\Gamma_0=|V_{cb}|^2G_F^2 m_b^5/(192\pi^3)$, $x=m_{\ep}^2/m_b^2$ and
$\tau_i(x)$ is given by:
\begin{eqnarray}
\tau_0(x)&=&x^2\log(1/x)
+{1\over 60} (1-x)
(3-27x-47x^2+13x^3-2x^4)
\nonumber\\
\tau_1(x)&=&
2x^2(2x+3)\log(1/x)+{1\over 6}(1-x)(1-11x-47x^2-3x^3)
\nonumber\\
\tau_2(x)&=&
-x(3x+2)\log(1/x)+{1\over 12}(1-x)(3+47x+11x^2-x^3)
\label{EQnine}
\end{eqnarray}
The value of $\Gamma_0/\Gamma_B$ 
where $\Gamma_B$ is the total width of the $B$, 
may be ascertained, at leading order, by
factoring the phase space effects of the $c$-quark into the semi-leptonic
branching ratio so that if $x_c=(m_c/m_b)^2$ then for $l=e,\mu$:
%
\begin{equation}
\Gamma(\BAR B\to l \BAR \nu_l X_c)=
((1-8x_c+x_c^2)(1-x_c^2)+12x_c^2\log{1\over x_c})\Gamma_0
\end{equation}
Thus using $\Gamma(\BAR B\to l \BAR \nu_l X_c)/\Gamma_B=0.1$\cite{pdb}
we obtain $\Gamma_0/\Gamma_B\approx 0.2$.

Next to deduce $H$ we describe $\psi\to \gamma \ep$ by assuming that
the $\psi$ is a weakly bound state of $c\BAR c$ quarks\cite{appelquist} in
conjunction with 
the coupling in
Eq.~\ref{EQfive} [see Fig.~2]. 
Explicit calculations then show that the
amplitude is dominated by on-shell gluons leading to
\begin{equation}
m^2_{\ep} H^2\cos^2\te=
{\pi\alpha\over2 \alpha_s^2}
{\Gamma(\psi\to \gamma \ep)\over \Gamma(\psi\to e^+e^-)}
{(1-r)(1+r)^2\over r^3 \log^2 r}
\label{EQten}
\end{equation}
where $r=(m_{\ep}/m_\psi)^2$.

We thus arrive at $H(0,0,m_\ep^2)\sim 1.8 GeV^{-1}$ 
for $\alpha_s(m_\psi)=.25$ and
$\te=-17^\circ$\cite{isgur,pdb} and other data from \cite{pdb}. As mentioned
before, there is considerable uncertainty in the value of $\te$. However, due
to our normalization with respect to $\psi\to \gamma\ep$, the rate for
$b\to s g \ep$ is independent of the precise value of $\te$. This value of
$H$ is in
rough accord with the coupling which one expects from an anomalous QCD
contribution to the $g$-$g$-$\ep$ vertex\cite{schroeder,followup}.

%
%
%
%
%
%
%

Finally, in order to estimate the rate for $b\to \ep g s$ we need to know the
value of $F_1$ and $F_2$. $F_1$ is calculated 
at leading order, for instance, in
\cite{hou_soni} where $F_1\approx -.20$. We can infer that QCD corrections do
not effect this result greatly by considering next-to-leading order (NLO)
calculations of penguin operators in \cite{Grigjanis,palmer}. To do this we
expand in terms of their $O_i$ operators:
\begin{eqnarray}
c_4O_4+c_6O_6
&=&
(1/2)(c_6-c_4)(O_6-O_4)
+(1/6)(c_6+c_4)(O_3+O_5)
\nonumber\\
&+&(c_6+c_4) 
\left[(\BAR b_a\  \gamma^\mu L {\lambda^i_{ab}\over 2}\ s_b)
\sum_q   ( \BAR q_c\  \gamma_\mu {\lambda^i_{cd}\over 2}\  q_d)\right]
\end{eqnarray}
where $c_i$ are the appropriate Wilson Coefficients of $O_i$ 
\cite{palmer,buras_long}. We
note that the last term corresponds to a color octet exchange with the same
quantum numbers as a gluon.
Thus
we attribute it to the exchange of a single
effective gluon from a penguin inferring 
that $F_1=4(c_4+c_6)/g_s$ which gives a
result that agrees with the above to $O(20\%)$. For instance, using the leading
order calculation in \cite{buras_long}
with $\Lambda_{\BAR{ms}}^{(5)}=225MeV$ 
so $\alpha_s=.21$ we obtain 
$F_1=-.168$.
The quantity $F_2$ may be taken
directly from this calculation since $F_2=c_{8G}=.143$.

We can now readily estimate the branching ratio for $b\to s\ep g$. Using
$F_1=-.168$, 
$H\cos\te=1.7 GeV^{-1}$ and $\Gamma_0/\Gamma_B=.2$ and taking $m_b=4.8
GeV$ we get 
$1.9\times 10^{-3}$ 
if just the electric form factor is used while
the presence of the magnetic form factor increases this result by about 
50\%
yielding $Br(b\to s g \ep)\sim 2.8\times 10^{-3}$.


The results quoted in \cite{cleo_data} however contain an acceptance cut
designed to reduce the background from events with charmed mesons\cite{ds}.
This cut requires that the energy of the $\ep$, $E_\ep\geq 2.2 GeV$ which is
equivalent to the cut in the recoil mass $m_{rec}\leq 2.35 GeV$

In order to understand the $m_{rec}$ spectrum from a $B$-meson, one must factor
in the fermi motion of the $b$-quark with respect to the meson. For this
purpose, following previous works \cite{greub}, we assume that the momentum of
the b-quark has a probability distribution of the form
\begin{equation}
P
\propto
e^{- |p|^2/\beta^2}.
\label{EQtwelve}
\end{equation}
which is suggested by a harmonic-oscillator like wave function. Fits to the
semileptonic decay spectrum at CLEO\cite{cleobsy} suggest a value of
$\beta=0.287 GeV$; $m_b=4.87GeV$. In our results we consider $\beta=.15-.45
GeV$, though in most cases the branching ratios are not strongly dependent on
$\beta$. When the cut is imposed however, the fraction passing the cut is
strongly dependent on $m_b$. Following CLEO ($b\to s \gamma$)~\cite{cleobsy} we
vary $m_b$ over the range of $4.7$ to $4.9 GeV$. Using $\beta=0.3$ GeV we
find that for $m_b=4.7$ GeV
 the branching fraction which passes the cut is $6.7\times 10^{-4}$ while
 if $m_b=4.9GeV$ then the corresponding branching fraction 
is $9.8\times 10^{-4}$. Furthermore the proportion of events passing the cut
varies from $\zeta_{cut}(E_\ep)\approx 25-33\%$.

Thus, taking $m_b=4.8 GeV$ and identifying $b\to s g \ep$ as the dominant
contribution\cite{followup} to the inclusive process we find that after
including the experimental cuts:
\begin{equation}
Br(B\to \ep X)\approx 8.2\times 10^{-4}
\ \ \ \ {\rm (including\ cuts)}
\label{EQfourteen}
\end{equation}
which is in rough accord with the experimental result given in Eq.~\ref{EQone}.
If we include a 150 MeV uncertainty in $m_b$ and 0.04 in $\alpha_s$ then in
this model the above number varies by about 30\%. Fig.~3 shows the $m_{rec}$
distribution for $\beta=0.15$, $0.3$ and $0.45GeV$, $m_b=4.8 GeV$ as well as
the case $\beta=0.3GeV$ for $m_b=4.6 GeV$. As can be seen, for a fixed $m_b$,
there is some variation in these spectra with $\beta$ only in the region
$m_{rec}\leq1.5 GeV$\cite{cleo_data}. The reason for the large dependence on
$m_b$ is readily apparent since as $m_b$ decreases the curve shifts to the
right and the proportion of events falling below the $m_{rec}$ cut
corresponding to $E_{\ep}=2.2 GeV$ rapidly drops.

Although there are several uncertain parameters in the calculation $(\alpha_s,$
$m_b, \beta\dots)$ it is quite remarkable that an
approximate constant form factor $H$
leads to a rate (Eq.~(\ref{EQfourteen})) that roughly agrees with experiment
(Eq.~(\ref{EQone})) as one of the gluons is considerably off-shell, $\langle
q^2_1\rangle\sim{}$10 GeV$^2$. 
This behavior of the form factor may be an indication that 
it is dominated by the presence of gluonic states in the two gluon channel.
The large $\ep$ signal offers a unique possibility, in any case, to search for
such states through the modes, for example,
\begin{eqnarray}
{\rm Glueball}\to \ep + h^++h^- 
\label{EQfifteen}
\end{eqnarray}
where $h=\pi$ or $K$. Thus it may be worthwhile to closely
study the invariant mass distribution of $(q_{\ep} + q_{h^+} + q_{h^-}
)^2$ in the data sample.

\section{Contribution from Charmonia}

Charmonia contributions (See Fig. 4) can be subdivided into three categories:
\begin{enumerate}

\item The most important of this class of contributions is from
$b \rightarrow \etac + s$ followed by $\etac \rightarrow \ep + X$.
(See Fig.~4a).
The experimental acceptance cut plays a very important role here. 
It essentially eliminates all contributions from a multi-body
state X and only a single particle or resonance (X) remains viable. 
Important
examples of X are: $\sigma$,$\eta$, $\omega$, $\ep\dots$~\cite{sigma}.
We find
$\zeta_{cut} (m_X \sim 2m_{\pi}) \sim 20 \%$, 
$\zeta_{cut} (m_X \sim m_{\eta}) \sim 15 \%$,     
$\zeta_{cut} (m_X \sim m_{\omega}) \sim 10 \%$ and  
$\zeta_{cut} (m_X \sim m_{\phi}) \sim 1.5 \%$.
From the measured branching ratios of $\etac$ into such states 
we estimate that their sum, 
$\sum_i Br(\etac \rightarrow \ep X_i) \sim 10\%$.
Using $Br(B \rightarrow \etac + X_s) \sim 7.0 \times 10^{-3}$ \cite{pdb,desh}, 
we thus see that, from this mechanism:
\begin{eqnarray}
Br(B \rightarrow \ep + X_s) &=& Br(B \rightarrow \eta_c +X_s) 
\times 
\sum_i Br(\etac \rightarrow \ep + X_i) \times \zeta_{cut}
\nonumber\\
&\sim& 1.1 \times 10^{-4}. 
\label{EQsixteen}
\end{eqnarray}
%
%
%

%
%
\item Next we consider $b \rightarrow \etac^* (\rightarrow \ep) + s$.
(See Fig. 4b). We view this contribution to arise through a mixing of
$\etac$ and $\ep$. Such a mixing is predominantly driven by the
two gluon intermediate state. It is very difficult
to see why this mixing angle should 
be bigger than a degree\cite{followup}.
Thus $Br(b \rightarrow \etac^*(\rightarrow \ep) + s) = \sin^2 \theta_{\etac}
\times Br(b \rightarrow \eta_c + s) \le 3 \times 10^{-6}$ and is
negligible.    
\item  Finally there is also the possibility that $b \rightarrow \psi + s$
followed by $\psi \rightarrow \ep + X$.  [See Fig.~4a] Again, the acceptance
cut 
effectively requires that X be a single particle state. Thus possible
examples of X are $\gamma$, $\omega$, $\phi$ etc. The corresponding branching
ratios are expected to be very small, each is  $\leq 0.5\%$ \cite{pdb},
suggesting 
that $\sum_i Br(\psi \rightarrow \ep + X_i) \sim 2\%$. We thus
estimate from $b\to\psi+s$ the resulting contribution is  $Br(B \rightarrow
\ep + X_s) \sim 3 \times 10^{-5}$.

\end{enumerate} 

Thus the main charmonia contribution is via $B \rightarrow \etac + X_s$, $\etac
\rightarrow  \ep + X$ with an effective 
$Br(B \rightarrow \ep + X_s) \sim 1.1 \times 10^{-4}$ 
(including cuts), 
which is at most 20\% of the observed central value. 

Experimentally, the charmonia contributions coming via cascade decays
of $\psi$ and or $\etac$ can, in principal, be separated by studying
the invariant mass distributions of the expected final states,
such as $\ep + \eta (\ep, \phi, \omega, \gamma, \pi \pi, KK etc)$,
which should cluster around $m_{\etac}$ and/or $m_{\psi}$.
  
Note also that while the $\etac-\ep$ mixing is expected to make a negligible
contribution it tends to have a distinctive $m_{rec}$ distribution 
which is quasi-two-body; it peaks around $m_{rec} \sim 1.5GeV$
and diminishes rapidly either side of this region.

\section{Other Mechanisms}

In passing  we note that there are other mechanisms\cite{followup} for making
$\ep$ in $B$ decays:
\begin{enumerate}

\item There is a $b\to s\BAR ss$ penguin process where the $\BAR s$-quark from
the gluon may pair with one of the $s$-quarks as shown in Fig.~5a.

\item There is a $b\to s\BAR uu$ (or $s\BAR dd$) penguin process where the
light quarks pair as shown in Fig.~5b.

\item There is a $b\to s u\BAR u$ (and also $b\to s d\BAR d$) penguin process
where the $u$($d$) can pair with the spectator $\BAR u$($\BAR d$) to make the
$\ep$ as shown in Fig.~5c.

\item There is a $b\to s gg$ penguin sub-process followed by $gg\to \ep$ shown
in Fig.~5d.

\item The $\ep$ may also be produced through a $b\to s u \BAR u $ tree graph
shown in Fig.~5e.

\end{enumerate}

Using the NLO effective Hamiltonian \cite{palmer} and vacuum saturation we can
estimate  the contributions of Fig.~5a and Fig.~5b. This is straightforward
for all of the operators except for the contributions of $O_5$ and $O_6$. In
these cases we need to know
\begin{equation}
G_{\eta_1}(q_i) =\sqrt{3}<0|\BAR q_i \gamma^5 q_i|\eta_1>
\end{equation}

Due to the presence of the anomalous contribution there is no direct way
to  deduce  the value of $G_{\eta_1}$ but a reasonable
estimate is perhaps that $G_{\eta_1}=G_\pi$ where
\begin{equation}
G_\pi
= <0|\BAR u \gamma^5 d|\pi^+>
\approx m_\pi^2 f_\pi/(m_u+m_d).
\end{equation}
Using $m_u+m_d\approx 15 MeV$, 
Fig.~5a together with Fig.~5b give $\sim 7.5\times 10^{-6}$.

We estimate the three-body penguin (Fig.~5c) to contribute about 
$2\times 10^{-5}$ which gives 
$1\times 10^{-5}$ with the cut implemented. In
this estimate, we have used as normalization, the CLEO central value
for $Br(B^0\to \pi^-\ell^+ \nu_\ell) = (1.8\pm.4\pm.3\pm.2) \times 10^{-4}$
\cite{cleo_data}.

Simma and Wyler \cite{simma_wyler} consider gluball production from $b\to s
gg$. Their calculation can be adapted to the $\ep$ (Fig.~5d) yielding about
$1\times 10^{-5}$.

The $b\to s u \BAR u$ tree graph contribution (Fig.~5e) may be estimated to be
about $5\times 10^{-6}$ 
which becomes about $2.6\times10^{-6}$ with cuts.
This contribution is only to $B^+$; the $B^0$ decay via the tree is color
suppressed. Thus when one averages over $B^0$ and $B^\pm$ 
this is effectively reduced by a factor of about 2.

Thus, for $b\to\ep+X_s$  
all of these processes are found to be appreciably less than the
production through the anomaly coupling (Fig.~1) and indeed not even competitive
with the charmonia contribution, Eq.~(\ref{EQsixteen}). We estimate that these
subdominant processes will not alter the main contribution by more than about
$10\%$; therefore, for now we can safely ignore them \cite{followup}.

Inclusive $\eta$ production, $B\to \eta + X_s$, will in general proceed through
the processes which we have discussed above for the $\ep$. The production
through the gluon mechanism of Fig.~1 is suppressed by approximately
$\tan^2(\te)$ (up to small additional corrections due to the difference in
mass). Thus if $\te=-17^\circ$ then the total branching ratio with this
mechanism is 
$2.6\times 10^{-4}$ while with a cut of $E_\eta>2.2GeV$ the
branching fraction is 
$8\times 10^{-5}$.

Next we consider the charmonia contribution to $\eta$. From Ref.~\cite{pdb} 
we see that in $\eta_c, \psi$ decays, 
the branching ratios of final states
with $\eta$ are roughly the same as those with $\ep$\cite{etacut}. 
This leads us to
suggest that, from charmonia, $Br(B\to\eta+X_s)\sim 7.3\times10^{-5}$.

The mechanisms Fig.~5a and 5b for the $\eta$ give $1.3\times 10^{-5}$. The
kinematics are such that virtually all of the events will pass the cut.
The process shown in Fig.~5c gives a total branching fraction
of about $3.6\times 10^{-5}$ 
of which about $1.3\times 10^{-5}$ passes the cut. The
tree process gives 
$\approx 3\times 10^{-6}$ (with cuts) but only from $B^\pm$
decays thus 
effectively reducing this by a factor of 2 when 
$B^\pm$ and $B^0$ are averaged.

It is also interesting to consider the situation when one replaces the $b\to s$
penguin with a $b\to d$ penguin and thus contemplates the inclusive process
$b\to \ep X_d$.

With the exception of the tree graph (Fig.~5e)
and the charmonium, all of the processes considered
above will scale with $r_{ds}^2=|V_{td}/V_{ts}|^2\sim O(\lambda^2)$ where
$\lambda = \sin\theta_c \simeq .22$ while the charmonium processes will scale
like $\lambda^2$.

The tree graph, on the other hand, leads to  
$\Gamma(b\to d\BAR u u )/\Gamma(b\to
s\BAR uu)\approx 1/\lambda^2$ so the contribution of this process to the
branching ratio of $b\to \ep X_d$ is about $5.4\times 10^{-5}$ (including
cuts). Therefore this
may be expected to provide the dominant contribution to $B^+\to \ep +
X_d$. For $B^0$ this process is color suppressed. So from the tree,
averaging over the two, $B\to\ep +X_d \sim 2.7\times 10^{-5}$
(with cuts).

\section{Summary and Outlook}

Table~\ref{tabone} summarizes the expected rates, for individual mechanisms, as
well as the total for the four final states of interest: $\ep
X_s$, $\ep X_d$, $\eta X_s$ and $\eta X_d$. For the
$\ep X_s$, the branching ratio is about 
$3.5\times 10^{-3}$ and
is about 
$2.5\times 10^{-4}$, 
$1\times 10^{-3}$ 
and
$1.4\times 10^{-4}$ 
respectively
for the other three, assuming for simplicity,
$|V_{td}/V_{ts}|^2 = 0.05$. The current CLEO acceptance cuts
[Eq.~(\ref{EQone})] affect the anomalous contribution (Fig.~1) more
severely (about a factor of 3 to 3.5) than the tree contribution (about
a factor of two). As indicated in the Table the anomalous graph is the
dominant mechanism contributing perhaps about 
86\% of 
the total for $\ep X_s$. It contributes about 
55\%,
44\% and 
10\% 
to $\ep X_d$, $\eta X_s$ and $\eta X_d$ respectively.

An especially notable feature of 
the numbers in the
Table is that the rates 
with cuts
for
the other modes 
($\eta X_d$, $\ep X_d$ and $\eta X_s$)
are also predicted
to be quite large, i.e. $(0.4\ {\rm to}\  1.8) \times 10^{-4}$,
although none is as high as the $\ep X_s$. Their searches 
are eagerly awaited as comparison with experiment will 
be very instructive. 

In closing, it is important to note that the $b\to s \ep g$ 
mechanism should be regarded as
the dominant contributor to the multiparticle inclusive process $B\to \ep +
X_s$, where $X_s$ is a kaon together with at least one pion. It is expected
\cite{followup} to
make a negligible contribution to the exclusive two body mode $(B\to \ep + K)$.
We recall that experimentally $B\to \ep + K$ is seen \cite{cleo_data} to be
about 10\% of the inclusive signal $B\to \ep +X_s$. To
facilitate  contact with theoretical models, it may be better if the two
body signal is not  included in the inclusive rate, Eq.~\ref{EQone}.

It is extremely interesting that the $\ep$ becomes such an effective probe of
the $b\to s g^*$ process. It has long been known that these charmless final
states have a very large branching ratio \cite{hou_soni}. The $\ep$ could
therefore become a very powerful tool for gluonia searches and for illuminating
our understanding of the 
penguin dynamics,  and, in particular for the quest for direct CP violation
as the same final states are expected to be rich in CP asymmetries\cite{bander}.

\bigskip

\noindent {\bf Acknowledgements}: The authors would like to acknowledge useful
discussion with Jim Alexander, 
Bruce Behrens, Tom Browder, Jos\'e Goity,
Nathan Isgur, 
Alex Kagan,
Peter Kim, Bill Marciano, Ron Poling, Laura Reina,
Jon Rosner and Sheldon
Stone. This research was supported in part by the U.S. DOE contracts 
DC-AC05-84ER40150 (CEBAF), DE-AC-76CH00016 (BNL).

\newpage

\begin{center}
{\large\bf Figure Captions}
\end{center}

Figure 1: The main mechanism for the inclusive $\ep$ production:  $b\to s
g^*$ graph followed by $g^*\to g \ep$ where the black 
circle denotes the anomalous $gg\ep$ coupling.

\medskip

Figure 2: $\psi\to \gamma gg$ followed by $gg\to \ep$ via the anomalous $gg\to
\ep$ vertex.

\medskip

Figure 3: Recoil mass distribution for $B\to X_s \ep s$ from the sub-process
$b\to g \ep s$. The events falling to the left of the vertical line pass the
$E_{\ep}>2.2GeV$ cut. For $m_b=4.8$ GeV, the solid curve shows the distribution
for $\beta=0.3GeV$ 
while in the region $m_{rec}<1.6 GeV$ the distributions for $\beta=0.15 GeV$
(dashed) and $\beta=0.45 GeV$ (dotted) are also shown. For $m_b=4.6$ GeV
and $\beta=0.3$ GeV the $m_{rec}$ distribution is shown by the dot-dashed
curve. Each of the above curves is
normalized so that its integral is unity.

\medskip

Figure 4: Charmonia contributions: 
a) $b\to\psi,\eta_c+s$ followed by
$\psi,\eta_c\to\ep+X$. b) $b\to\eta^\ast_c (\to\ep) +s$; 
{\bf X} denotes
$\eta_c\leftrightarrow\ep$ mixing.
\medskip

Figure 5: Feynman diagrams for the sub-dominant contributions to $B\to \ep +
X_s$. The four quark penguin operators (i.e. $O_3-O_6$) are represented by
hexagons. The diagrams shown are: (a) the $b\to s\BAR s s $ process where
the $\BAR s$-quarks may combine with either of the $s$-quarks to form an
$\ep$; (b) the $b\to s \BAR u u$ or $s\BAR d d$ process where the
$u\BAR u$ ($d\BAR d$) forms the $\ep$; (c) A  typical 
penguin graph producing two
gluons that fuse to form a $\ep$; (d) a $b\to s \BAR u u$ or $s\BAR d
d$ process where the light quark combines with the spectator; (e) the $b\to s
u\BAR u$ tree process.

\newpage

\begin{table}[h]
\caption{Contributions from different mechanisms to the
$Br(B\to\ep (\eta) + X_{s,d})$ in units of $10^{-5}$. The
number from each graph is given along with the total, without cuts. The
effect of CLEO acceptance cuts on the
total is also indicated. Note
$r^2_0=20 |V_{td}/V_{ts}|^2$.\label{tabone}}
\begin{center}
\begin{tabular}{|l|c|c|c|c|c|}
\hline
Mechanism & Type & $\ep X_s$ & $\ep X_d$ & $\eta X_s$ & $\eta
X_d$  \\ \hline
Fig.~1 & Anomaly &  
$280$ 
& $14r^2_0$  
&  26 
& $ 1.3r^2_0$ 
\\ \hline 
Fig.~4 & Charmonia 
& $70$ 
& $3.4$ 
& $70$  
& $3.4$ 
\\ \hline
Fig.~(5a + 5b) & penguin-4q-op 
&  0.75 
& $0.04r^2_0$ 
& 1.3 
& $0.07 r^2_0$ \\ \hline 
Fig.~5c & penguin-4q-op 
& 2.1  
&  $0.1r^2_0$ 
& 3.6 & 
$0.18r^2_0$ \\ \hline 
Fig.~5d & penguin-2g-op 
& 1.0 
&  $0.05 r^2_0$ 
& 0.1 
& $0.005r^2_0$ \\ \hline 
Fig.~5e & tree  
& 0.25 
&  5.2
& 0.4 
& 8.6 \\ \hline 
Total (no cuts) & & 
$ 354$ 
& $14r^2_0+8.6$ 
& 101 
& $1.6r^2_0 +12$ \\ 
\hline
Total (with cuts) & 
& $96$ 
& $4.2r^2_0+3.2$ 
& 17.6 
& $.54r^2_0+3.3$ 
\\ \hline 
${{\rm Anomalous}\over{\rm total}}$
(with cuts)
& 
& $\sim 86$\% 
& $\sim 55$\%$^{a)}$ 
& $\sim 44$\% 
& $\sim 10$\%$^{a)}$ \\ \hline
\end{tabular}
\medskip
\end{center}
\noindent $a)$ assuming $r^2_0=1$.
\end{table}

%
%
%
%
%
%
%
%
%
%
\newpage
\begin{figure}[h]
\epsfysize 4.5in
\mbox{\epsfbox{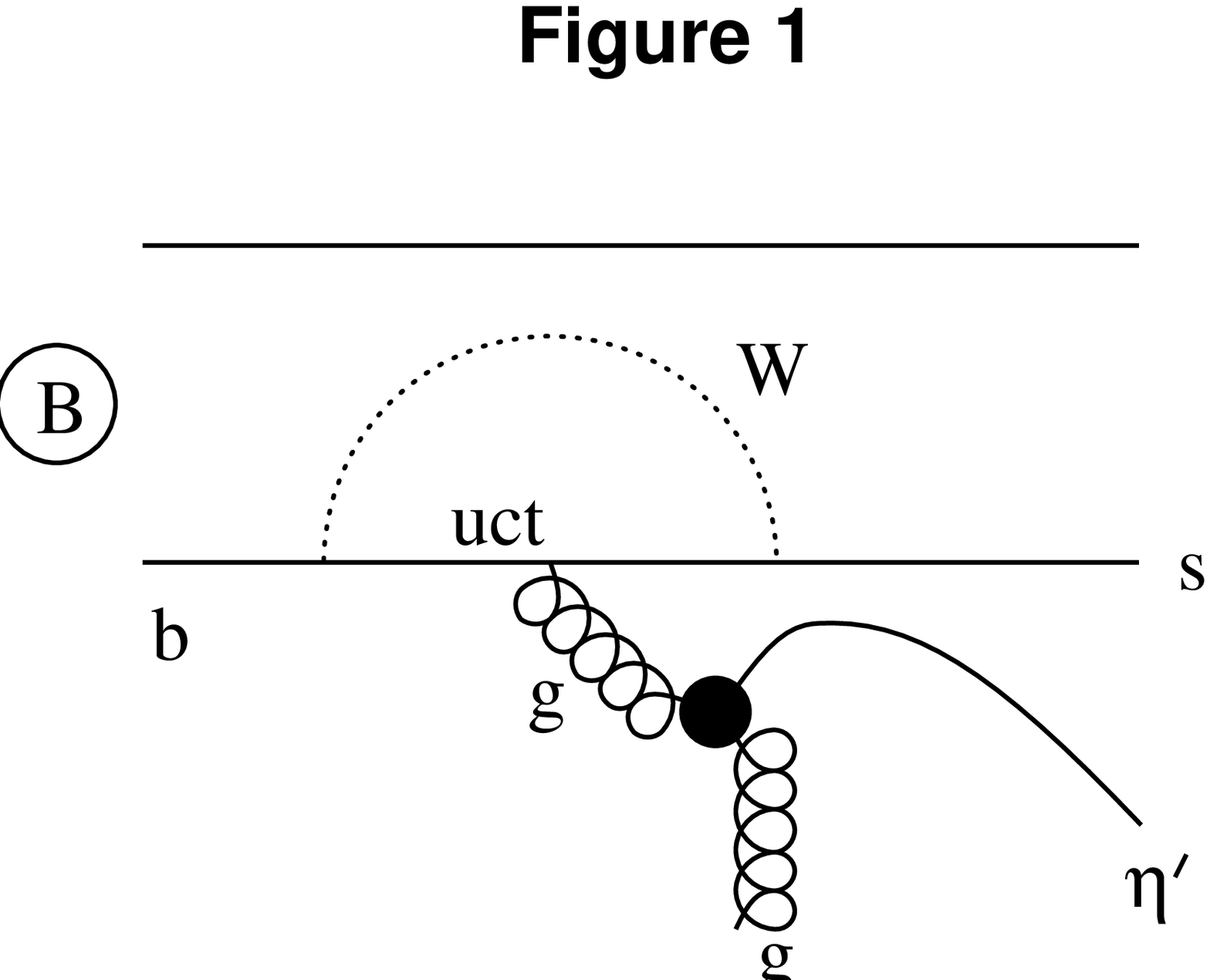}}
\end{figure}
\newpage
\begin{figure}[h]
\hspace*{-0.5in}
\epsfxsize 6.7in
\mbox{\epsfbox{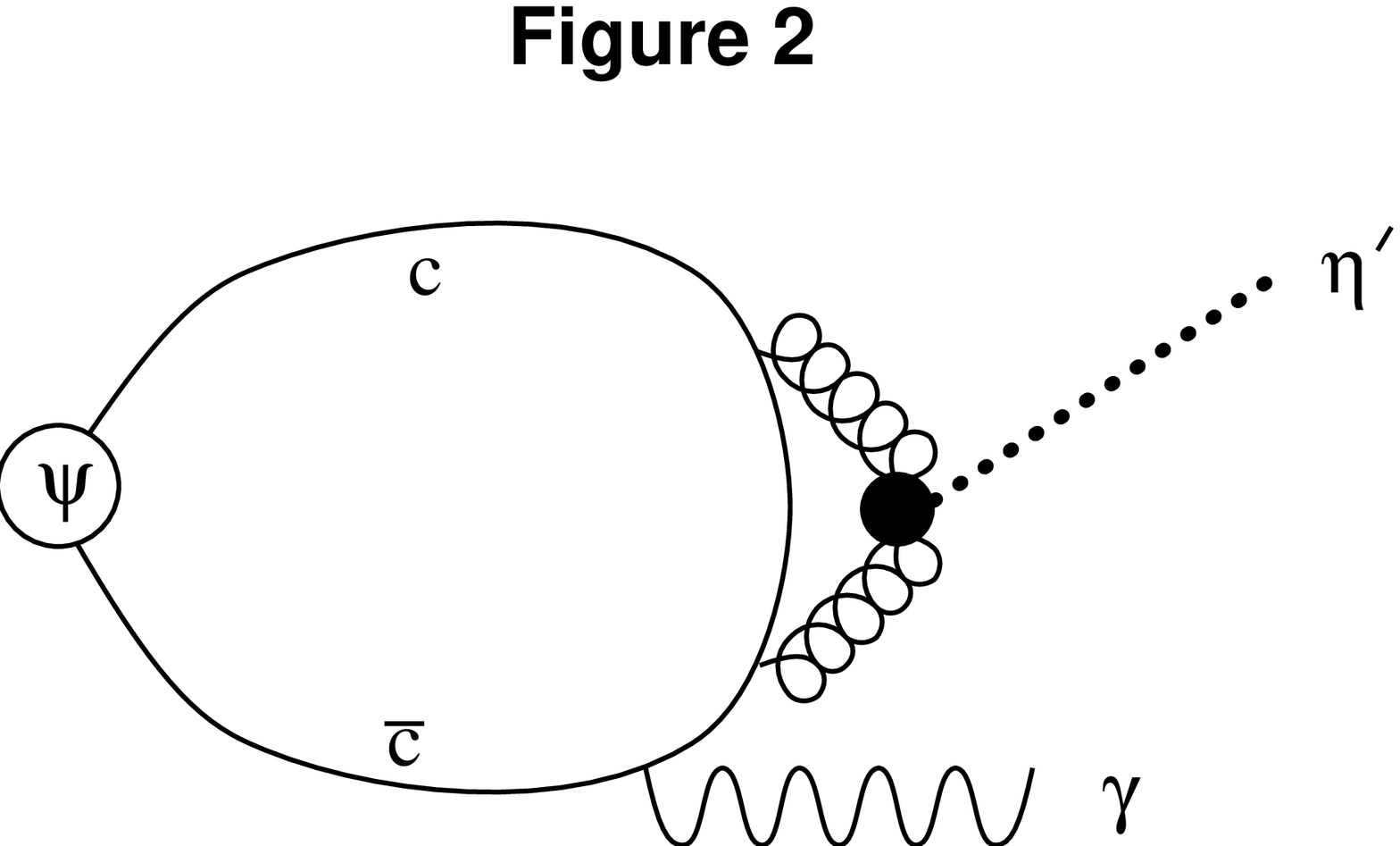}}
\end{figure}
\newpage
\begin{figure}[h]
\epsfysize 7in
\mbox{\epsfbox{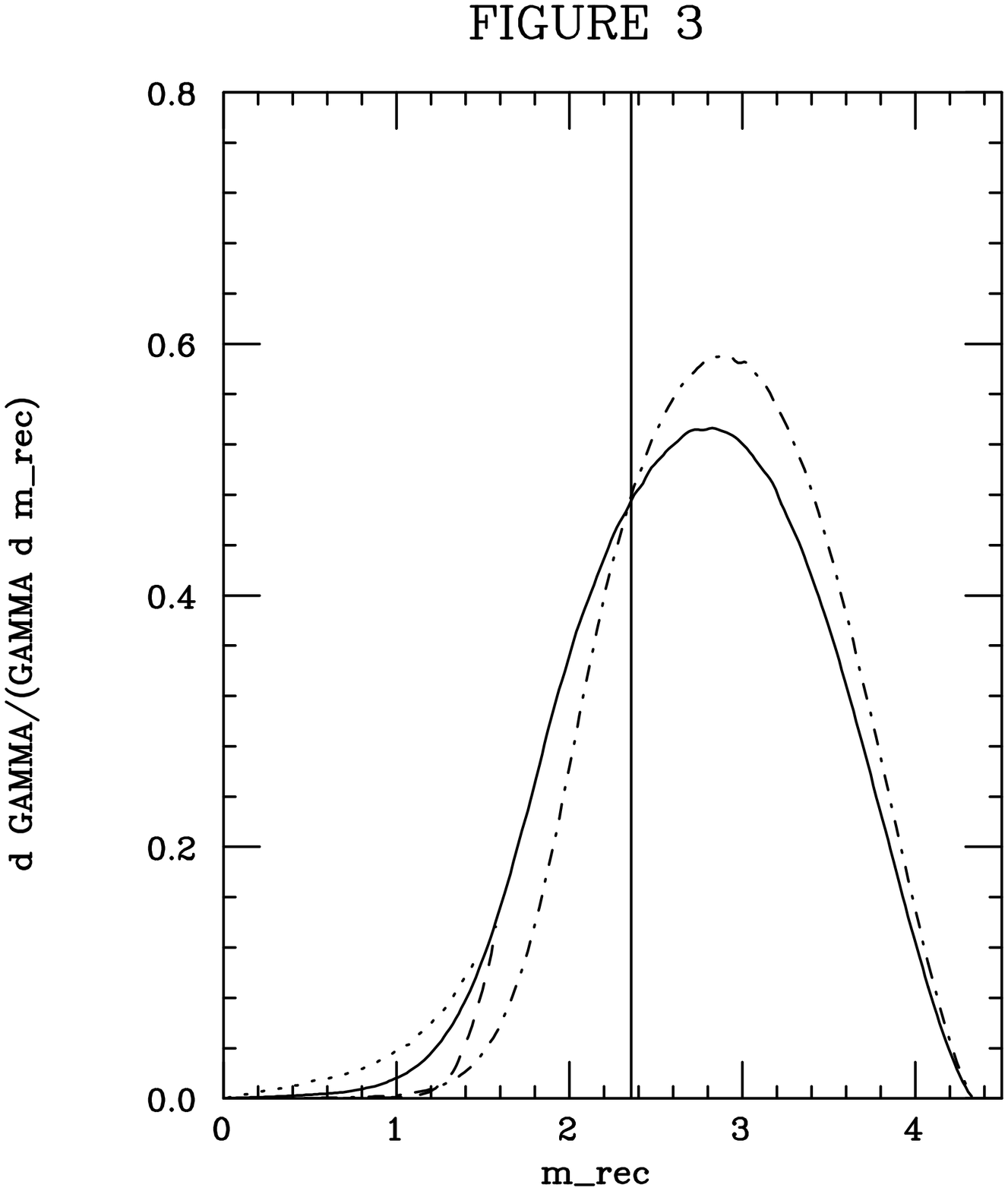}}
\end{figure}
\newpage
\begin{figure}[h]
\hspace*{-0.5in}
\epsfxsize 6.0 in
\mbox{\epsfbox{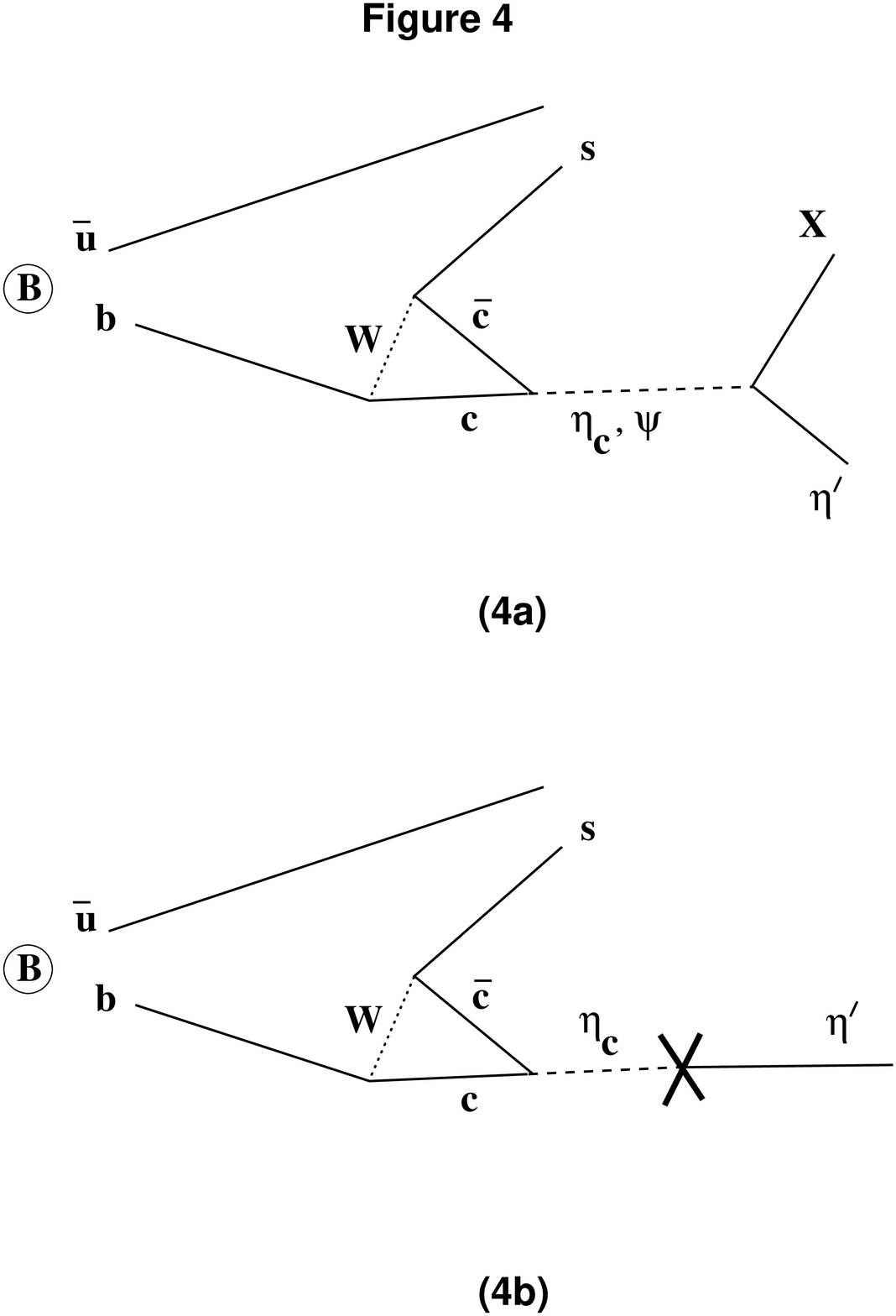}}
\end{figure}
\newpage
\begin{figure}[h]
\hspace*{-0.5in}
\vspace*{-1.0in}
\epsfxsize 5.5 in
\mbox{\epsfbox{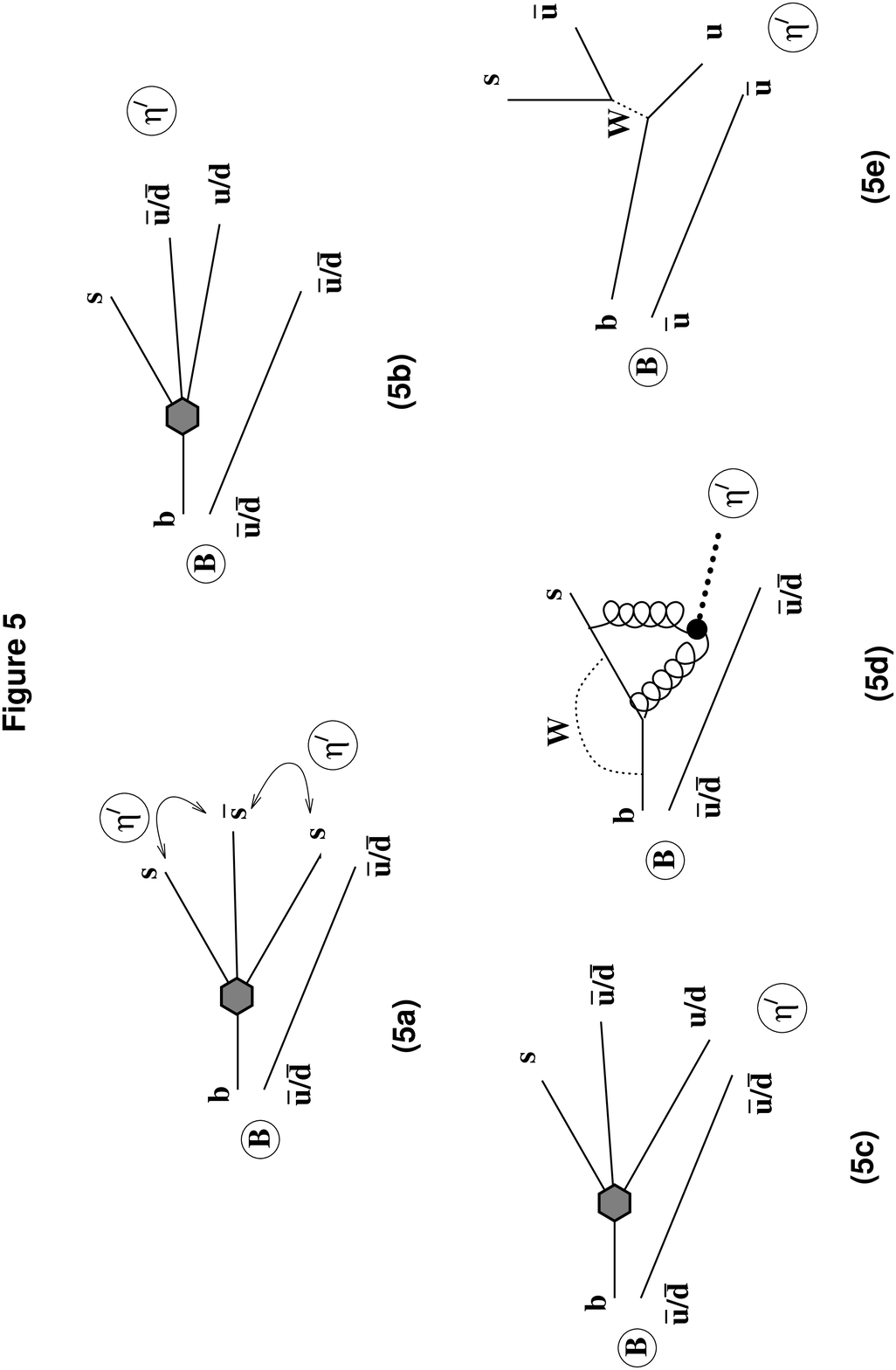}}
\end{figure}

\begin{thebibliography}{99}


\bibitem{cleo_data} P.~Kim, [CLEO] Talk at FCNC 1997, Santa Monica, CA (Feb
1997); See also the  [CLEO] talks by B. Behrens and by R. Poling at $B$ Physics
and CP Violation, Waikiki, HI (March 1997).


\bibitem{hou_soni} W.~S.~Hou, A.~Soni and H.~Steger, Phys.~Rev.~Lett.~{\bf59},
1521 (1987); W.~S.~Hou, Nucl.~Phys.~B{\bf308}, 561 (1988).

\bibitem{Grigjanis} R.~Grigjanis, P.~O'Donnell, M.~Sutherland and H.~Navelet,
Phys.\ Lett.\ B{\bf224}, 209 (1989); B.~Grinstein, R.~Springer and M.~Wise,
Phys.\ Lett.\ 
B{\bf202}, 138 (1988); A.~Buras, M.~Jamin, E.~Lautenbacher and P.~Weisz,
Nucl.\  Phys.\
B{\bf370}, 69 (1992); M.~Ciuchini, E.~Franco, G.~Martinelli, L.~Reina and
L.~Silvestrini Phys.\ Lett.\ B{\bf344}, 137 (1994).

\bibitem{schroeder} See e.g. D.~V.~Schroeder and M.~Peskin {\it An
Introduction to Quantum Field Theory}, Addison-Wesley (1995); C.~Itzykson and
J.~B.~Zuber, {\it Quantum Field Theory}, McGraw-Hill (1980).


\bibitem{derujula} A.~DeRujula, H.~Georgi and S.~Glashow, Phys.\ Rev.\ D{\bf12},
147 (1975); N.~Isgur, Phys.\ Rev.\ D{\bf13} (1976); E.~Witten, Nucl.\ Phys.\
B{\bf156}, 269 (1979).


\bibitem{pdb} Particle Data Group, Phys.\ Rev.\ D{\bf54}, 1 (1996).

\bibitem{rosner} J.~Rosner, Phys.\ Rev.\ D{\bf27}, 1101 (1983).


\bibitem{Tytgat} P.~Ball, J.-M.~Fr\'ere and M.~Tytgat, Phys.\ Lett.\ B{\bf365},
367 (1996) and references therein.


\bibitem{appelquist} T.~Appelquist and H.~D.~Politzer, Phys.\ Rev.\ D{\bf12},
1404 (1975).

\bibitem{isgur} N. Isgur in Ref.~\cite{derujula}.

\bibitem{followup} For further details see D.~Atwood and A.~Soni, in
preparation.


\bibitem{palmer} G.~Kramer, R.~Palmer and H.~Simma, Z. Phys.\ C{\bf66}, 429
(1995). 


\bibitem{buras_long}
G.~Buchalla, A.~J.~Buras and M.~E.~Lautenbacher, Rev. Mod. Phys.
{\bf 68}, 1125 (1996).



\bibitem{ds} A large background is expected from decays of the type 
$B \to D_s + D + X$ followed by $D_s \to \ep + Y$ (e.g. $D_s \to \ep +
\rho$). The experimental acceptance cut in Eq. (1) eliminates most
of these.    


\bibitem{greub} See A.~Ali and C.~Greub, Phys.\ Lett.\ B{\bf259}, 182 (1991) and
references therein; N.~Isgur, D.~Scora, B.~Grinstein and M.~Wise, Phys.\ Rev.\
D{\bf39}, 799 (1989); G.~Altarelli {\it et. al}., Nucl.\ Phys.\ B{\bf208},
365 (1982). 


\bibitem{cleobsy} M.~S.~Alam {\it et. al}. (CLEO Collab.), Phys.\ Rev.\ Lett.\
{\bf74}, 2885 (1995).

\bibitem{sigma} We are using the notation $\sigma$ to mean a $0^{++}$,
broad resonance in the $2\pi$ channel, with mass${}\sim400$--1200 MeV and
width 600--1000 MeV [See Ref.~6].


\bibitem{desh} N. G. Deshpande and J. Trampetic, Phys.\ Lett.\ B{\bf339}, 
270(1994). 



\bibitem{simma_wyler} H.~Simma and D.~Wyler, 
Nucl.\ Phys.\ B{\bf344}, 253 (1990).


\bibitem{etacut} In the case of the charmonia mechanism giving rise to
$\eta$, 
the cuts tend to
reduce the signal more than in the case of $\ep$.
Thus
$\zeta_{cut} (m_X \sim 2m_{\pi}) \sim   14 \%$,
$\zeta_{cut} (m_X \sim m_{\eta}) \sim   10 \%$,
$\zeta_{cut} (m_X \sim m_{\omega}) \sim  5 \%$ and
$\zeta_{cut} (m_X \sim m_{\phi}) \sim     .25 \%$.
We will therefore use an average cut of $10\%$ for the $\eta$ compared
to $15\%$ for the $\ep$.

\bibitem{bander} M.~Bander, D.~Silverman and A.~Soni, Phys.\ Rev.\ Lett.\
{\bf43}, 242 
(1979); N.G. Deshpande and A. Soni, Proc.\ Snowmas '86, p. 58; 
J.~M.~Gerard and W.~S.~Hou , Phys.\ Rev.\ D{\bf43}, 2909 (1991);
H.~Simma, G.~Eilam and D.~Wyler, 
Nucl.\ Phys.\ B{\bf352}, 367 (1991); G.~Kramer,
R. Palmer and H. Simma (ref.
\cite{palmer})
L.~L.~Chau, H.~Y.~Cheng, W.~K.~Sze, B.~Tseng and H.~Yao, Phys. Rev. 
{\bf D45}, 3143 (1992) and references therein.


\end{thebibliography}
\end{document}